\documentclass{llncs}
\usepackage{llncsdoc}

\usepackage{amsmath}

\usepackage{epsfig}
\usepackage{graphicx}
\usepackage{epsfig, graphics}
\usepackage{amssymb,amstext}
\usepackage{colortbl}
\usepackage{eurosym}
 \usepackage{amsfonts}
\usepackage{vaucanson-g}

\usepackage{psfig}
\usepackage{float}

\newcommand{\out}[1]{}

\begin{document}

\title{Temporal Support of Regular Expressions in Sequential Pattern Mining}

\author{Leticia I. G\'omez \inst{1} \and Bart Kuijpers \inst{2} \and  Alejandro Vaisman \inst{2}}
\mainmatter

\authorrunning{G\'omez, Kuijpers, Vaisman}

\institute{Instituto Tecn\'ologico de Buenos Aires\\
\email{lgomez@itba.edu.ar}\\
 \and
 University of Hasselt and \\
 Transnational University of Limburg\\
\email{{alejandro.vaisman,bart.kuijpers}@uhasselt.be}}

\maketitle

\begin{abstract}
Classic algorithms for sequential pattern discovery,
return  all  frequent sequences
present in a database. Since, in general, only a few ones
 are interesting from a user's point of view,
 languages based on
 regular expressions (RE) have been  proposed
  to restrict frequent
 sequences to the ones that satisfy
user-specified constraints.
 Although  the support of a sequence is computed
 as the number of data-sequences satisfying a pattern with
  respect to the total number
 of data-sequences in the database, once regular expressions come into play,
 new approaches to the concept of support  are needed. For example,
 users may be interested in computing the support of the RE as a whole,
  in addition to the  one  of a particular pattern.
 Also, when the items are frequently updated,  the
 traditional way of counting support in sequential pattern
  mining may lead to   incorrect (or, at least incomplete), conclusions. For
    example, if we are looking  for the support of the sequence  A.B, where A
    and B are two items such that A was created \emph{after} B, all sequences in the
     database that were completed \emph{before} A was created, can never produce
     a match. Therefore, accounting for them would underestimate the
      support of the  sequence A.B.
  The problem gets more involved if we are interested in categorical sequential patterns.
  In light of the above, in this paper we propose to revise the classic  notion of support
 in sequential pattern mining,  introducing the concept of \emph{temporal
 support of regular expressions}, intuitively defined as
 the number of sequences satisfying a target pattern, out of the total number of
 sequences that \emph{could have possibly} matched such pattern, where the pattern is
 defined as a RE over complex items (i.e., not only item identifiers,
 but also attributes and functions).
 \end{abstract}


 \section{Introduction}
\label{sec:introduction}

 Traditional sequential patterns algorithms are founded on the assumption  that  items in databases are \emph{static}, and that they existed throughout the whole lifespan of the world modeled by the database. There are many real-world situations where sequential pattern mining (SPM) is usually applied, and where these assumptions are not valid any more. In these situations, items are created or deleted dynamically. Further, if we are interested in categorical SPM, we need to deal with complex items, i.e., items described by attributes (or even functions over attributes). These attributes are also usually subject to change. Consider for example SPM in \emph{trajectory databases.} For many applications, we may be interested in trajectory patterns involving restaurants, hotels, gas stations. The features that characterize these places may change over time, and even many of them could have not existed when some of the trajectories under analysis occurred. This may also occur in the context of the World Wide Web, where Web pages are frequently added or deleted. Ntoulas \emph{et al.}~\cite{Ntoulas01} collected snapshots over 155 web sites, during one year, once a week. They concluded  that new pages are created at the rate of 8\% per week, and only 20\% of the pages available at  one instant will be accessible after one year.  Thus, there ia a high  frequency of creation and deletion of Web pages. Moreover, they found that  the link structure of the Web is more dynamic that the page content.

We introduce the problem through a   Web usage mining example.
Data Mining techniques have been applied for  discovering interaction patterns of WWW users. Typically,
 this mining is performed over the URLs visited during a session, recorded in a  Web server log. In this way,
 the interests and behavioral patterns  of Web users  can be  studied.
 Figure \ref{fig:runningexample} depicts a portion of a   (simplified) Web log.
  In classic SPM, the support of a sequence $S$ is defined as the fraction of sessions that support $S$.
   Thus,  all sessions are considered as having the \emph{same probability} to support a given sequence.
  For example, the support of the sequence CBC, counted in the classical way, would be  66\%, since
  CBC  is present in two of the three sessions. Analogously, the support of the sequence CB would also be 66\%.
  We may ask  would have happened if not all these Web pages existed all the time. The question is:
   would it be   realistic to count support in the usual way?  More precisely, would it be reasonable  to ignore the
   evolution of the items (URLs) across time? We
    discuss these issues in this paper.

\begin{figure}[t]
\centering
 \psfig{figure=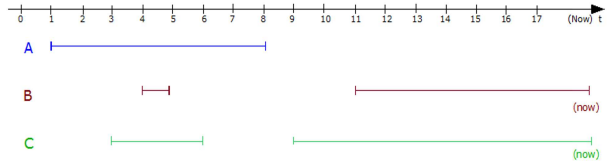,width=8cm,height=2.2cm}
        \caption{Evolution of three URLs A, B, and C}
\label{fig:temporalmining}
\end{figure}

 When a Web page is visited during a  session, it is often the case where  a user clicks a
nonexisting link or a link that has been removed. Figure \ref{fig:temporalmining} shows how
URLs A, B, and C  in Figure \ref{fig:runningexample},
  have evolved, and the time intervals  when each  URL has been available.
 We can see that URL A was available during the interval [1, 8], URL B  in  intervals [4, 5] and [11, \emph{now}],
 and  URL C, during [3, 6] and [9, \emph{now}]. (We use the term \emph{now} to refer to the current time instant).
 We now analyze the support of the sequence CBC. During session s2, we can see that URL C did not
 exist at  t=8,  when the user clicked URL A. Thus, session s2 did not have the possibility of producing
 a sequence that finishes with the URL C. Sessions s1 and s3, instead, support the
 sequence CBC. Then, ignoring the evolution of these URLs,
  the support of sequence CBC would be 66\%, but,  if we do not count session s2,
  we would obtain  a support of 100\% for this sequence.
Analogously, if we  compute the support of the sequence of CB taking into account the \emph{availability} of the items during each session, we can see that  s1 and s3 support this sequence, but  s2 does not. However,  C was available during session s2, when
 the user clicked URLs A (t=3) and  B (t=4)  (actually, it existed in the interval  [3, 6]). Thus, the user
 could have produced the sequence CB, although she decided to follow a different  path.  Session s2 must then be counted   for computing the support of the sequence CB, which would be 66\%.

The example above gives the intuition of the ideas that we  discuss and formalize in this paper:  the support of a sequence depends on the counting method, and when items evolve over time, new definitions of support are needed.  Instead of considering \emph{all sequences} in the database in the same way, we  propose to account for the fact that some of these sequences  could have never been produced due to the temporal unavailability of some of the items in them.

\begin{figure}[t]
  \begin{scriptsize}
\begin{center}
\begin{tabular}{|c|c|}
  \hline
  session ID & interaction \\
  \hline
  s1 & \begin{tabular}{c|c}
        time & URL \\
         \hline
         t=1 & A \\
         t=3 & C \\
         t=5 & B \\
         t=6 & C \\
       \end{tabular}
          \\
  \hline
  s2 & \begin{tabular}{c|c}
        time & URL \\
         \hline
         t=3 & A \\
         t=4 & B \\
         t=8 & A \\
       \end{tabular}
          \\
  \hline
  s3 & \begin{tabular}{c|c}
        time & URL \\
         \hline
         t=5 & A \\
         t=10 & C \\
         t=20 & B \\
         t=23 & C \\
       \end{tabular}
          \\
  \hline
\end{tabular}
\end{center}
  \end{scriptsize}
\caption{Web user interaction}
\label{fig:runningexample}
\end{figure}


\subsection{Related Work}
\label{sec:relatedwork}

 Sequential Pattern discovery in databases has been studied for a long time. Classic algorithms~\cite{Agrawal95,Srikant96}  return  all  frequent sequences present in a database. However, more  often than not, only a few ones are interesting  from a user's point of view. Thus,  post-processing tasks are required in order to discard  uninteresting   sequences. To avoid this drawback,  languages based on regular expressions (RE) were  proposed  to restrict frequent sequences to the ones that satisfy user-specified constraints.
 Garofalakis \emph{et al.}~\cite{Garofalakis01,Garofalakis02} address this problem by  pruning the candidate patterns obtained during the mining process by adding  user-specified constraints in the form of  regular expressions over items. The algorithm  returns only the frequent patterns that satisfy these regular expressions.
    Toroslu and Kantarcioglu \cite{Toroslu01} limit the number of sequences to be found through  a parameter called
   \emph{repetition support}. The idea  consists in detecting cyclically repeating patterns. The parameter
  specifies the minimum number of repetitions of the patterns within each data-sequence. Thus, the algorithm finds
  frequent sequences with at least minimum support and a cyclic repeating pattern.

 Recently, the data mining community started to discuss   new notions of support in SPM, that account for changes   of the  items database across time. Although  this problem
 has already been addressed for Association Rule mining, where the concept of \emph{temporal support} has already been introduced~\cite{Lee01,Li03,Tansel07}, this has  been overlooked in SPM. To the best of our knowledge,
 the works we  comment below are the only ones partially addressing the issue.

Masseglia \emph{et al.} \cite{Masseglia01}, and Parthasarathy {et al.}\cite{Parthasarathy01}, study the so-called \emph{incremental sequential pattern mining} problem. This problem arises when items are appended to a database. They focus on designing efficient algorithms in order to avoid   re-scanning  the entire database when new items appear. They address the  addition of items to existing transactions, and
 the addition of new transactions. In the absence of new transactions, the
 previously computed frequent patterns will still be frequent in the new database, and the  problem  consists in detecting the occurrences of \emph{new} frequent patterns. In the presence of new transactions, however,
 old frequent patterns may or may not be frequent in the \emph{incremental database}.
   Recently, Huang \textit{et al.} \cite{Huang01}  address the problem of detecting frequent patterns \emph{valid during a defined  period of interest,} called POI. For example, if new items appear, and no new transactions were generated, old frequent sequences would  still be frequent.

\subsection{Contributions}
\label{sec:contribution}

 In addition to the problem of item evolution and availability commented above,
 we believe that other scenarios have been
 overlooked so far. For example, when regular expressions (from now on, RE) are
 used to prune non-interesting patterns, we may ask ourselves if a  user would
 be interested not only in the support of a sequence, but in the support of an RE as a whole.
 Let us analyze a simple example. The expression $(A|B).C$ is satisfied by
 sequences like A.C or B.C. Even though the semantics of this RE
 suggests that both of them are
 equally interesting to the user, if neither of them verifies a  minimum support (although
 together they do), they would  not be retrieved.
 The problem gets more involved if we are interested in categorical sequential patterns, i.e.,
 patterns like \emph{Science.Sports}, where \emph{Science}
 and \emph{Sports} are, for instance, categories of Web pages in an ontology
(in SPIRIT~\cite{Garofalakis01,Garofalakis02},
  the alphabet of the REs is composed only of
 item identifiers).

 In light of the above,  we propose to revise, in different ways, the classic
 notion of support for sequential pattern mining.
 We introduce the concept of \emph{temporal
 support of regular expressions}, intuitively defined as
 the number of sequences satisfying a target pattern, out of the total number of
 sequences that \emph{could have possibly} matched such pattern, where the pattern is
 defined as a RE over complex items.
  We first introduce the data model (Section \ref{sec:datamodel}), then we
 present and discuss a theoretical framework for this novel notion of support, and an RE-based
 language (Sections \ref{sec:simplelanguage} and \ref{sec:querylanguage}).  We conclude in
  Section \ref{sec:conclu}.

 \section{Data Model}
\label{sec:datamodel}

Depending on the application domain, the items to be mined can be characterized by different attributes.
Throughout the paper we refer to an example where  each Web page is  characterized by the following attributes:  (a) \emph{catName}, which represents the name of the category of the item\footnote{Although in our running example we have only one category as an instance of \emph{catName}, there are other applications where this is not the case. For example, in  a trajectory database application analyzing  tourist itineraries, items could be categorized as hotels,  restaurants, or tourist attractions, to name a few ones. Each one of them could be characterized by different attributes. For instance, the kind of food offered by a restaurant could be an attribute of the category \emph{restaurant}.}; (b) \emph{keyword}, which summarizes the page contents; (c)  \emph{filter},  specifying  a list of URLs that cannot appear together with the URL of the item. Finally, \emph{ID} is a distinguished, mandatory attribute, in this case containing the URL that
 univocally identifies   a Web page.
For each category there are  \emph{occurrences}. In our example, we work with three URLs, for simplicity referred to as A, B, and C. We denote \emph{set of instances} a set of occurrences of a collection of categories. The items to be mined are events defined over some category occurrence at some instant. These items are stored in a so-called Table of Items (ToI). In Figure \ref{fig:items} we show a ToI for our running example.


\begin{figure}[t]
  \begin{scriptsize}
    \centering
  \begin{tabular}{|c|l|}
    \hline
    OID & Items \\
    \hline
    $S_{1}$ &
        \begin{tabular}{l}
        [(t, `08/04/2008~14:05'), (ID,`B'),\\
        (catName,`WebPage'),(filter,`A,C'),(keyword,`Game')]\\
        ...\\
        $[$(t,`08/08/2008~17:10'), (ID,`A'),\\
        (catName,`WebPage'),(filter,`'),(keyword,`Computer')]\\
        \end{tabular} \\
    \hline
    $S_{2}$ &
        \begin{tabular}{l}
        [(t,`08/03/2008~11:00'), (ID,`C'),\\
        (catName,`WebPage'),(filter,`A'),(keyword,`Computer')]\\
        ...\\
        $[$(t,`08/19/2008~09:00'), (ID,`A'), \\
         (catName,`WebPage'),(filter,`'),(keyword,`Computer')]\\
        \end{tabular} \\
    \hline
  \end{tabular}
  \caption{An instance of a Table of Items (ToI)}
  \label{fig:items}
\end{scriptsize}
\end{figure}

\subsection{Introducing Temporality}

In many real-world applications, assuming that the values of attributes for a category occurrence do not change (or even that a category occurrence spans over the complete lifespan of the dataset) could not be realistic.
 Thus, we   introduce the time dimension into our data model. We do this
 in the usual way, namely, timestamping   category occurrences. We assume that the category schema is constant across time, i.e., the attributes of a category are
 the same throughout the lifespan of the category.

\begin{definition} {\rm[Category Schema]}
\label{def:catsch}
We have  a set of attribute names \textbf{A}, and a set of identifier names \textbf{I}.
Each attribute $\mathit{a} \in \textbf{A}$ is associated with a set of values in $\mathit{dom(a)},$
and each identifier $\mathit{ID} \in \textbf{I}$ is associated with a set of values in $\mathit{dom(ID)}.$

\rm A \emph{category schema} S is a  tuple
$(\mathit{ID},\mathit{A}),$
where $\mathit{ID} \in \textbf{I}$ is a distinguished attribute denoted \emph{identifier},
 and
$\mathit{A}$ is a set of attributes in $\textbf{A}$. Without loss of generality, and for simplicity,
in what follows we  consider the set $\mathit{A}$  ordered. Thus, S has the form
$[\mathit{ID}, attr_{1}, ... , attr_{n}].$ \qed
\end{definition}

\begin{example}
In  our running example we have only one category, representing Web pages with schema $[$$\mathit{ID}$, $\mathit{catName}$, $\mathit{filter}$, $\mathit{keyword}$$]$.\qed
\end{example}

 We  consider the time as a new sort (domain) in our model.
  Toman~\cite{Toman96}, showed the equivalence between
 \emph{abstract} and \emph{concrete} temporal databases. The former are \emph{point-based}
 structures, independent from the actual implementation of the database. The latter contains
 efficient interval-based encodings of the former.
 The author also showed that there is an efficient  translation from abstract to concrete temporal
 databases.   Formally, if
 T is a set, and $<$ a discrete linear order without endpoints on T,  the structure
 $T_P=(T,<)$ is the \emph{Point-based Temporal Domain}.
The elements in the carrier  of $T$ model the individual time instants, and the linear
 order $<$ models the succession of time. We consider
the set $T$ to be \textbf{N} (standing for the natural numbers).
We can map individual time instants t $\in$ \textbf{N} to  calendar instants, assuming a reference point and a granularity. For example, if the reference point is January 1, 1970 00:00 GMT, and  granularity ``minute'',
 t=1440 represents 1440 minutes from that date, i.e., January 2, 1970 00:00 GMT. In what follows we use calendar time, and granularity ``minute''.
In temporal databases, the concepts  of \emph{valid} and \emph{transaction} times refer, respectively, to the instants when data is valid in the real world, and when data is recorded in the database~\cite{Tansel93}.  We assume \emph{valid time} support in this paper for the categories, and \emph{transaction time} support for the items (see Definitions \ref{def:catocc} and \ref{def:item} below).

\begin{definition}{\rm[Category Occurrence]}
\label{def:catocc}
Given a category schema $\mathit{S},$ a \emph{category occurrence} for $\mathit{S}$
is the tuple $[\langle ID, id\rangle,\mathit{P},t],$ where  $\mathit{ID}$ is the ID  attribute of
Definition \ref{def:catsch} above,  $\mathit{id} \in dom(ID),$
 $\mathit{P}$
is the structure
 $[\langle attr_{1},v_{1}\rangle $..., $\langle attr_{n},v_{n} \rangle]$,
 $t$  is a point in the temporal domain $T_P,$ and: (a)
 $attr_{i}= \mathit{A}(i)$ in $\mathit{S}$ (remember that $\mathit{A}$ is considered ordered);
 (b) $v_{i} \in dom(attr_{i}), \forall i, i=1..n;$
 (c) All the occurrences of the same category have the same set of  attributes, at any given time;
 (d) At any instant $t$, the pair $\langle \mathit{ID},t\rangle $ is unique for
 a category occurrence, meaning that no two occurrences of the same category can have
 the same value for $\mathit{ID}$ at the same time;
 (e) $t$ is the time instant when  the information in the category occurrence is valid.
\end{definition}

\begin{definition} {\rm [Category Instance]}
\label{def:catinst}
A set of occurrences of the same category is denoted a
\emph{category instance.}  We extend the
 fourth condition in Definition \ref{def:catocc} to hold for the whole set:
 no two occurrences of categories in the set can have the
  same value for $\mathit{ID}$ at the same instant $t$ (in other words, the pair $(ID,t)$ is unique
   for the whole instance).\qed
\end{definition}

\begin{remark}
In what follows, for clarity, we  assume that
$attr_{1}$  stands for  $\mathit{ID}$. Thus,
a category occurrence is the set of pairs
 $[\langle$  $attr_{1}$,$v_{1}$$\rangle$,..., ... , $\langle$ $attr_{n}$,$v_{n}$$\rangle$, t$]$. \qed
\end{remark}

Since  point-based and interval-based representations are equivalent,
in this paper we work with  the  latter. One of the reasons for this is that in an actual implementation,
we work with intervals.
In our encoding, an event is  represented by an interval whose endpoints are the same.
We need to define this encoding in a precise way. The following definition states the condition
that a set of tuples must satisfy in order to belong to the same group.

\begin{definition}{\rm [Interval Encoding]}
\label{def:partition}
Let $G$ be a time granularity, and $g$ a time unit for  $G$ (e.g., one minute).
Given a set of $k  \geq 0$ category occurrences, $[\langle  attr_{1},v_{1}\rangle ,\ldots, \langle attr_{n},v_{n}\rangle,t_{1}]$,
 $[\langle attr_{1},v_{1} \rangle , \ldots, \langle attr_{n},v_{n}\rangle,t_{2}],\ldots,[\langle  attr_{1},v_{1}\rangle \ldots$\\
   $\langle attr_{n}, v_{n}\rangle, t_{k}],$  if $\forall~i, i=1..{k-1},$ it holds that $t_{i+1} =  t_{i}+g,$  we
  encode all these occurrences in a single tuple
$[\langle attr_{1},v_{1}\rangle,\ldots,\langle attr_{n},v_{n}\rangle,[t_{1},t_{k}]]$.
 \qed
 \end{definition}

\begin{figure}[t]
  \begin{tiny}
  \centering
        \begin{tabular}{|l|lllll|}
        \hline
        $eco_{A1}$ & [(ID,`A'), & (catName,`WebPage'), & (filter,`'), & (keyword,`Books'), & [`11/29/2007 15:45', `11/29/2007 17:10'] ] \\
        $eco_{A2}$ & [(ID,`A'), & (catName,`WebPage'), & (filter,`P'), & (keyword,`Computers'), & [`11/29/2007 18:00', `11/29/2007 19:30']] \\
        $eco_{C1}$ & [(ID,`C'), & (catName,`WebPage'), & (filter,`A'), & (keyword,`Books'),  & [`11/29/2007 16:00', `11/29/2007 16:45']] \\
        $eco_{C2}$ & [(ID,`C'), & (catName,`WebPage'), & (filter,`'), & (keyword,`Games'), & [`11/29/2007 18:00', `11/29/2007 18:50']] \\
        $eco_{C3}$ & [(ID,`C'), & (catName,`WebPage'), & (filter,`M'), & (keyword,`Games'), & [`11/29/2007 19:30', `Now']] \\
        $eco_{M1}$ & [(ID,`M'), & (catName,`WebPage'), & (filter,`'), & (keyword,`Games'), & [`11/29/2007 18:50', `11/29/2007 20:00']] \\
        $eco_{P1}$ & [(ID,`P'), & (catName,`WebPage'), & (filter,`'), & (keyword,`Computers'), & [`11/29/2007 16:45', `11/29/2007 18:50']] \\
        $eco_{P2}$ & [(ID,`P'), & (catName,`WebPage'), & (filter,`A'), & (keyword,`Computers'), & [`11/29/2007 18:51', `Now']] \\
        \hline
        \end{tabular}
\end{tiny}
\centerline{\psfig{figure=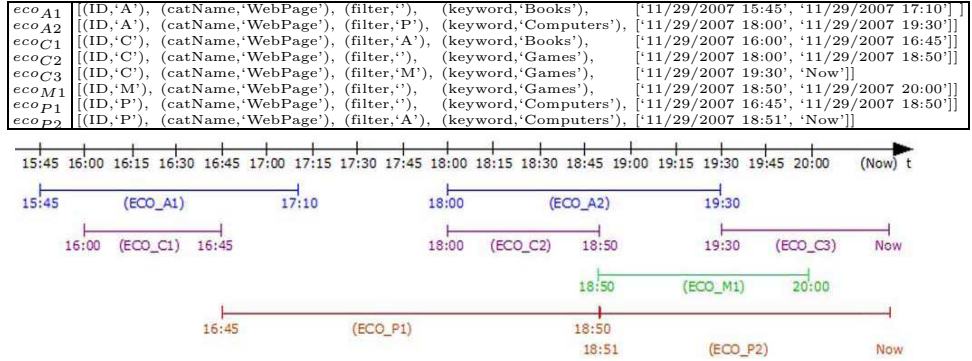,width=12cm,height=3cm}}
\caption{Encoded Category Occurrences for the running example (granularity ``minute'').}
\label{fig:eco-example}
\end{figure}

\begin{example}
Figure \ref{fig:temporalocurrencesexample} shows a  set of (point-based) temporal category occurrences for the Web page category in our running example.  Figure \ref{fig:eco-example}  shows the corresponding  interval-encoded representation (see below for  details).

 Encoding a set of tuples requires these tuples to be consecutive over the granularity selected. Thus, if the granularity is ``minute'', the tuples  $[(ID, \mbox{`A'}),$ $(keyword, \mbox{`computer'}), (filter, \mbox{`'}), \mbox{`12/12/2000~12:31'})]$,  and   $[ (ID, \mbox{`A'}), (keyword,$ \\ $\mbox{`computer'}), (filter, \mbox{`'}),$ $\mbox{`12/12/2000~12:33'})],$  cannot be included together in the same group, since there is a two-minute gap between them. They must be encoded into two intervals. \qed
\end{example}

\begin{figure*}[t]
  \begin{scriptsize}
 \begin{center}
\begin{tabular}{|c|c|}
\hline
Category & Instance \\
\hline
Web Page &
        \begin{tabular}{lllll}
        \hline
        $[$ (ID,`A'), & (catName,`WebPage'), & (filter,`'), & (keyword,`Books'), & `11/29/2007 15:45' $]$ \\
        $[$ (ID,`A'), & (catName,`WebPage'), & (filter,`'), & (keyword,`Books'), & `11/29/2007 15:46' $]$ \\
        ... & & & & \\
        $[$ (ID,`A'), & (catName,`WebPage'), & (filter,`'), & (keyword,`Books'), & `11/29/2007 17:10' $]$ \\
        $[$ (ID,`A'), & (catName,`WebPage'), & (filter,`P'), & (keyword,`Computers'), & `11/29/2007 18:00' $]$ \\
        $[$ (ID,`A'), & (catName,`WebPage'), & (filter,`P'), & (keyword,`Computers'), & `11/29/2007 18:01' $]$ \\
        ... & & & & \\
        $[$ (ID,`A'), & (catName,`WebPage'), & (filter,`P'), & (keyword,`Computers'), & `11/29/2007 19:30' $]$ \\
        ... & & & & \\
        $[$ (ID,`M'), & (catName,`WebPage'), & (filter,`'), & (keyword,`Games'), & `11/29/2007 18:50' $]$ \\
        $[$ (ID,`M'), & (catName,`WebPage'), & (filter,`'), & (keyword,`Games'), & `11/29/2007 18:51' $]$ \\
        ... & & & & \\
        $[$ (ID,`M'), & (catName,`WebPage'), & (filter,`'), & (keyword,`Games'), & `11/29/2007 20:00' $]$ \\
        ... & & & & \\
        \end{tabular} \\
\hline
\end{tabular}
 \end{center}
 \end{scriptsize}
\caption{Category occurrences, granularity ``minute'' (Point-based).}
\label{fig:temporalocurrencesexample}
\end{figure*}



\begin{definition} {\rm [Encoded Category Occurrence]}
\label{def:eco}
Given a category instance $\mathcal{C}$ with time  granularity G,  and  a partition $\mathcal{P}$ of $\mathcal{C}$ such that the
number of sets $p_i \in \mathcal{P}$ is minimal. Each set $p_i$ is obtained encoding the occurrences in $\mathcal{C}$ as in Definition \ref{def:partition}, i.e.,  each $p_i$ contains a set of tuples that can be encoded into a  single tuple. Thus, associated to  $p_i$ there is a tuple $t_{p_i}=(\langle ID, id \rangle, \langle attr_1, v_1 \rangle,\ldots\langle attr_n, v_n \rangle, t_v, t_e ),$ where (a)
   ID, $attr_1$, ..., $attr_n$ are the attributes of the   occurrences in $p_i$; (b)
   id, $v_1$, .... $v_n$ are the values for the attributes in (a);
   (c)  $t_s$ is the smallest t of the occurrences in $p_i$;
   (d) $t_e$ is the largest $t$ of the  occurrences in $p_i$.
 We denote  $t_{p_i}$ an encoded category occurrence (ECO) of the set of occurrences in $p_i.$
 Given an ECO $e_i$ we denote $\mathit{Interval}(e_i)$ its associated interval $[t_s, t_e].$\qed
\end{definition}

\begin{example}
\label{ex:tempcatocur}
 Figure \ref{fig:eco-example} shows a set of eight ECOs encoding the category instance
of Figure \ref{fig:temporalocurrencesexample}.
The Web page with ID=`P' (not included in Figure~\ref{fig:temporalocurrencesexample}) had no filter when it was created until November 29th, 2007 at 18:50, when the attribute \emph{filter} was updated.
Also, the Web page with ID=`A' has changed: attribute  \emph{keyword} was updated from `Books' to `Computers', and  also \emph{filter} was updated. Note that  there is an interval when this   page was not available. After these  changes, the page was set off-line at 7:30PM on November 29th, 2007.\qed

\end{example}

Adding a time instant to an ECO, produces
an \textbf{item}.

\begin{definition} [Item]
\label{def:item}
Let $e_o=(\langle ID,v \rangle,\langle attr_{1},v_{1} \rangle,$ $\ldots\langle attr_{n},v_{n}\rangle,t_s, t_e)$ be an ECO
for some category instance.  An \emph{item} I associated to $e_o$
 is the  set: $(\langle t,v_{t} \rangle,\langle ID,$
 $v \rangle, \langle attr_{1},v_{1} \rangle,\ldots\langle attr_{n},v_{n}\rangle,t_s, t_e\rangle),$
 such that $v_{t}\in [t_s, t_e]$ holds. We denote  $t$ the \emph{transaction time} of the item.\qed
\end{definition}

\section{A Theory for Support Count}
\label{sec:simplelanguage}

 In Section \ref{sec:datamodel} we defined the formal data model we use in the remainder
 of the paper to build a theory that can help to provide insight into the notion of support.
 To begin with,
 over the elements introduced in Definition \ref{def:catsch}
 through \ref{def:item}, we build a  simple language, based on sequences of constraints,
 that we use later to elaborate the concept of support of regular expressions.
  In short, this language  expresses paths of constraints.
  We define the temporal support of these paths, denoted \emph{sequential expressions} (SE).
  SEs are at   the cornerstone of our theory. In the next section,
  we   define a regular language that produces SEs, and introduce the notion of \emph{temporal support
  of regular expressions}.

\begin{definition}{\rm [Terms]}
\label{def:terms}
There exist no others terms than the following ones:
  (a) Constant:  a literal enclosed by simple quotes;
  (b) Non temporal Attribute: an attribute in the category schema (e.g. filter, url).
   (c) Temporal Attribute: t, the temporal attribute of and item (see Definition \ref{def:item});
   (d) Function of n arguments: Let $f_n$ be a function symbol, the expression
    $f_n(\mathit{attribute}, \mbox{`ct1'}, \mbox{`ct2'}, ... ,\mbox{`ct}_{n-1}\mbox{'}),n \geq 1,$ is a function
    where the \emph{first} parameter is an attribute (temporal or non-temporal),
    and all the other ones are constants.\qed
\end{definition}

\begin{definition}{\rm [Atoms]}
\label{def:atom}
Let C, A and F be a set of constants, temporal and non temporal attributes and functions, respectively.
The expression term1 = term2 is an atom, where term1 $\in$ A $\cup$ F,  term2 $\in$ C, and  $\mbox{`}=\mbox{'}$
 is the equality symbol.\qed
\end{definition}

\begin{definition}{\rm [Formula]}
\label{def:formula}
We define recursively a formula by the following rules:
(a) An atom is a formula; (b) If F1 and F2 are formulas
 then F1 $\wedge$ F2 is a formula.
 (c) Nothing else is a formula.\qed
\end{definition}

\begin{definition}{\rm [Constraint and Formula of a Constraint]}
\label{def:constraint}
A \emph{constraint} is a formula enclosed in squared brackets.
Given a constraint $C=[\mathcal{F}]$, we denote $\mathcal{F(C)}$ the \emph{formula of} C.\qed
\end{definition}

\begin{definition}{\rm [Sequential Expression]}
\label{def:SE}
A \emph{Sequential Expression} (SE) of length n is an ordered list of n
sub-expressions $SE_1.SE_2....SE_n$, where each $SE_i$ is a constraint, $\forall i,i=1..n$ \qed
\end{definition}

\begin{example}
The sequential expression of length two $[ID=\mbox{`}A\mbox{'}~\land~ filter=\mbox{`}B,C\mbox{'}].[ID=\mbox{`}X\mbox{'}]$ is composed of  two constraints.
\end{example}

We  need to define some operations between intervals.
Given two intervals $I_i=[ts_i, te_i]$ and $I_j=[ts_j, te_j]$ we say that $I_i$ \emph{follows}
 $I_j$ if $ts_i$ $\geq$ $te_j$.
 Saying that an interval $I_i$ \emph{follows} another interval $I_j,$ is equivalent to say
 that $I_i$ is either \emph{after} $I_j$ or $I_i$ is \emph{met-by }$I_j$ in terms of
 Allen's Interval Algebra \cite{Allen01}.

\begin{example}
In Figure \ref{fig:eco-example} we can see that Interval($eco_{C3}$) \emph{follows}
Interval($eco_{A2}$) and Interval($eco_{C2}$). We can also see that Interval($eco_{C3}$) \emph{does not follow} \\
 Interval($eco_{M1}$).
\end{example}

\begin{definition}{\rm [Satisfability of a Constraint]}
\label{def:constraintsatisf}
Given a constraint C and an ECO E, we say that E \emph{satisfies} C if one of the following conditions hold:
  (a)   If $\mathcal{F}(C)$ is an atom of the form $\mathit{attr}=\mbox{`ct'}$ where $\mathit{attr}$ is an
        attribute in any of the category occurrences in E, $\mathit{`ct'}$ is a constant in $\mathit{dom(attr)},$ and the instantiation of $\mathit{attr}$ with its value in E, equals  $\mathit{`ct'}$.
  (b) If $\mathcal{F}(C)$ is an atom of the form
   $f_n(\mathit{attr}, \mbox{`ct1'}, \mbox{`ct2'},...,\mbox{`ct}_{n-1}\mbox{'})=\mbox{`ct'},$ where $\mathit{attr}$ is an
        attribute in any of the category occurrences in E, $\mathit{`ct'}$ is a constant in $\mathit{dom(attr)},$
        and the instantiation of $\mathit{attr}$ in $f_n$ with its value in E, makes the equality \emph{true}.
   (c) If $\mathcal{F}(C)$ is an atom of the form $t=\mbox{`ct'}$ where $t$ is a
   temporal attribute, $\mbox{`ct'}$ is a temporal constant in the temporal domain, with  granularity G, and $\mbox{`ct'} \in \mathit{Interval}(E).$
  (d) If $\mathcal{F}(C)$ is an atom of the form    $f_n(t, \mbox{`ct1'}, \mbox{`ct2'},...,\mbox{`ct}_{n-1}\mbox{'})=\mbox{`ct'},$ where $t$ is a temporal attribute, $\mbox{`ct'}$ is a temporal constant in the temporal domain with some granularity G, and $\exists t_u \in \mathit{Interval}(E)$ the equality is \emph{true.}
   (e) If $\mathcal{F}(C)$ is a formula  $F1 \wedge F2,$ and $F1$ and $F2$ are satisfied by E.

\end{definition}

\begin{definition}{\rm [Satisfability of  SE]}
\label{def:SEsatisfability}
Let  $EO=(EO_1, EO_2,...EO_n)$ be a list of ECOs
 such that
  $\forall~i,j,$ \\
  $i <j$  $\Rightarrow$ $\mathit{Interval}(E_i)~\mathit{does~not ~follow}~\mathit{Interval} (E_j)$. We denote $EO$ a \emph{t-ordered} list of ECOs.
A sequential expression SE=$SE_1.SE_2....SE_n$ is satisfied by $EO$
if  $EO_i$ satisfies $SE_i$, $\forall~i,i=1..n.$
We denote $\mathcal{S}_{L_{k}}(SE)$ the set composed of the n lists of   ECOs
that satisfy an SE of length $k$. \qed
\end{definition}

\begin{example}
 Let us analyze which ordered lists of ECOs in Figure \ref{fig:eco-example} satisfy the  SE
  $[\mathit{rollup}(t, \mbox{`}\mathit{hour}\mbox{'}, \mbox{`}\mathit{Time}\mbox{'})=\mbox{`}18 \mbox{'}].[\mathit{keyword}=\mbox{`}\mathit{Books}\mbox{'}].$
  Here, \emph{rollup} is the usual rollup function~\cite{Cabibbo97},
   that indicates how a member of an OLAP hierarchy is aggregated. The meaning is that the equality
   is \emph{true} when $t$ is instantiated with a value that, in the Time dimension, rolls up to the value `18' in the dimension level \emph{hour}. For example,
    $[\mathit{rollup}(\mbox{`11/29/2007 18:52'}, \mbox{`}\mathit{hour}\mbox{'}, \mbox{`}\mathit{Time}\mbox{'})=\mbox{`}18 \mbox{'}].$

The first constraint is satisfied by $eco_{A2}$, $eco_{C2}$, $eco_{M1}$, $eco_{P1}$ and $eco_{P2}$.
For all of them, there is a time instant within $Interval(eco_i)$ that verifies the rollup predicate.
The second constraint is satisfied by $eco_{A1}$ and $eco_{C1}$.
However, given the temporal order, the only list of ECOs that satisfy the SE is:
$L_1= \{eco_{P1}, eco_{A1}\}.$ In $L_1,$ $[\mbox{`11/29/2007 16:45'},\mbox{`12/29/2007 18:50'}]$ (the interval of $eco_{P1}$) does not follow  $[\mbox{`11/29/2007 15:45'},\mbox{`12/29/2007 17:10'}]$ (the interval of $eco_{A1}$). \qed
\end{example}

%


\begin{definition}{\rm [ToI and Normalized  ToI]}
\label{def:tableofitems}
 Let $\mathcal{I}$ be a finite set of items. A \emph{Table Of Items} (ToI) for $\mathcal{I}$
  is a table with schema  $\mathit{T}=(OID, Items)$, where  \emph{Items}  is the name of an attribute whose instances are items, and an
  instance of $\mathit{T}$ is a finite  set of tuples of the form $\langle O_{j}, i_{k}\rangle$ where $i_{k} \in \mathcal{I}$ is an item associated to the object $O_{j}.$ Moreover, given $\langle O_{j}, i_{k}\rangle$ and $\langle O_{j}, i_{m}\rangle$, two tuples corresponding to the same object, and $t_k$ and $t_m$ the transaction times of the items, then $t_k \neq t_m$ holds.
 A \emph{normalized ToI} is a database containing a  table with
 schema $(OID, t,  ID)$ (the  \emph{Normalized ToI}), and one table
  per category, each one with schema $(ID,attr_{1},...,attr_{n},t_s, t_e).$ \qed
\end{definition}

Figure \ref{fig:ntoi-example} shows an instance of a  normalized ToI where items are related to
  the category instances of Figure \ref{fig:eco-example}.
There are  three sessions (sequences), $Session_{1}$, $Session_{2}$ and $Session_{3}$, each one with
 an associated list of items. The three sessions clicked on  URL C, but only $Session_{1}$ would satisfy the constraint
 $[ID=\mbox{`C'} \land catName=\mbox{`Books'}]$ (see  Figure \ref{fig:eco-example}).

\begin{figure}[t]
\centering
  \begin{scriptsize}
  \begin{tabular}{|c|l|}
    \hline
    OID   & Items \\
    \hline
    $Session_{1}$ &
        \begin{tabular}{lrr}
       $[$(t,`11/29/2007 16:30'), (ID,`C') $]$ &
            \textcolor[rgb]{0.50,0.00,0.50}{$\rightsquigarrow$} &
            \textcolor[rgb]{0.50,0.00,0.50}{$eco_{C1}$} \\
       $[$(t,`11/29/2007 17:00'), (ID,`P') $]$ &
            \textcolor[rgb]{0.79,0.00,0.00}{$\rightsquigarrow$} &
            \textcolor[rgb]{0.79,0.00,0.00}{$eco_{P1}$} \\
       $[$(t,`11/29/2007 19:45'), (ID,`C'),$]$ &
            \textcolor[rgb]{0.50,0.00,0.50}{$\rightsquigarrow$} &
            \textcolor[rgb]{0.50,0.00,0.50}{$eco_{C3}$} \\
        \end{tabular} \\
    \hline
    $Session_{2}$ &
        \begin{tabular}{lrr}
        $[$(t,`11/29/2007 18:20'), (ID,`C')$]$ &
            \textcolor[rgb]{0.50,0.00,0.50}{$\rightsquigarrow$} &
            \textcolor[rgb]{0.50,0.00,0.50}{$eco_{C2}$} \\
        $[$(t,`11/29/2007 18:50'), (ID,`P')$]$ &
            \textcolor[rgb]{0.79,0.00,0.00}{$\rightsquigarrow$} &
            \textcolor[rgb]{0.79,0.00,0.00}{$eco_{P1}$} \\
        $[$(t,`11/29/2007 18:51'), (ID,`M')$]$ &
            \textcolor[rgb]{0.00,0.50,0.00}{$\rightsquigarrow$} &
            \textcolor[rgb]{0.00,0.50,0.00}{$eco_{M1}$} \\
        \end{tabular} \\
    \hline
        $Session_{3}$ &
        \begin{tabular}{lrr}
        $[$(t,`11/29/2007 19:31'), (ID,`C')$]$ &
            \textcolor[rgb]{0.50,0.00,0.50}{$\rightsquigarrow$} &
            \textcolor[rgb]{0.50,0.00,0.50}{$eco_{C3}$} \\
        $[$(t,`11/29/2007 19:32'), (ID,`M')$]$ &
            \textcolor[rgb]{0.00,0.50,0.00}{$\rightsquigarrow$} &
            \textcolor[rgb]{0.00,0.50,0.00}{$eco_{M1}$} \\
        $[$(t,`11/29/2007 20:00'), (ID,`C')$]$ &
            \textcolor[rgb]{0.50,0.00,0.50}{$\rightsquigarrow$} &
            \textcolor[rgb]{0.50,0.00,0.50}{$eco_{C3}$} \\
        \end{tabular} \\
    \hline
  \end{tabular}
  \caption{An instance of the \emph{Normalized ToI}}
  \label{fig:ntoi-example}
\end{scriptsize}
\end{figure}


\begin{definition}{\rm [Temporal Matching of a S.E]}
\label{def:temporalmatching}
Let SE be a sequential expression of length $k,$ and
 a \emph{normalized ToI} (from now on, nToI), with schema $(OID,t,ID).$
An  object identified by $OID_m$ \emph{temporally matches} SE, if there exist k tuples in nToI,
$\langle OID_m, t_1,ID_1\rangle, \langle OID_m,t_2,ID_2\rangle,\ldots,\langle OID_m, t_k, ID_k\rangle$, where
 for at least one  $L_p \in \mathcal{S}_{L_{k}}(SE)$, $L_p=(eco_1, eco_2,\ldots,eco_k)$,
$t_i \in \mathit{Interval}(eco_i),  \\
\forall~i=1..k.$
\qed
\end{definition}

\begin{example}
\label{ex:temporalmatch}
 Definition \ref{def:temporalmatching}  states that if there is a temporally ordered  sequence of $k$ items such that all of their transaction times fall within the intervals of the $k$ ECOs that satisfy the expression, then, we have a temporal match.

With the category occurrences of Figure \ref{fig:eco-example} and the instance of  nToI
depicted in Figure \ref{fig:ntoi-example}, we analyze the  sequential expression
 SE = $[ID=\mbox{`P'}].[\mathit{filter}=\mbox{`M'}].$
The ECOs that satisfy the first constraint are $eco_{P1}$ and $eco_{P2}$. The second constraint is satisfied by $eco_{C3}$. Thus, the lists that satisfy SE are $L_1= \{ eco_{P1}, eco_{C3}\}$ and $L_2= \{eco_{P2}, eco_{C3}\}$.
The object $\mathit{Session}_{1}$ temporally matches  SE, since  there exist two  different tuples in $\mathit{Session}_{1}$ whose transaction times  belong to $\mathit{Interval}(eco_{P1})$ and $\mathit{Interval}(eco_{C3})$, respectively.
With a similar analysis, $\mathit{Session}_{2}$  \emph{does not match} the SE. The ECO $eco_{C3}$ did not exist when the user in this session clicked the last two URLs.
Finally, $\mathit{Session}_{3}$ temporally matches SE, because  the transaction time of $[(t,\mbox{`11/29/2007 19:32'}), (\mbox{ID,`M'})]$ belongs to the interval of $eco_{P2},$ and the transaction time of $[(t,\mbox{`11/29/2007 20:00'}), (\mbox{ID,`C'})]$ belongs to the interval of $eco_{C3}.$ Intuitively, this means that
 the user of $\mathit{Session}_{3}$ could have chosen the URL with  ID=`P', which existed at the time
 she chose the URL with ID=`M'.\qed
\end{example}

 From Definition \ref{def:temporalmatching}, it follows that if a list of ECOs does
  not \emph{satisfy} a sequential expression SE,  no object in the  nToI can use this list to temporally match SE. Thus, given that the lists in
 $\mathcal{S}_{L_{k}}(SE)$ are computed over the category occurrences, which usually fit in main memory,
  unnecessary database scans can be avoided.

\begin{definition}{\rm[Temporal Satisfability of a Constraint]}
\label{def:atomsatisf}
Given a constraint C and  a normalized ToI,  with schema $(OID, t,  ID),$
 we say that  a tuple in nToI $\mu=\langle OID_m, t_m, ID_m \rangle$ \emph{temporally satisfies} C
 if at least one of the following conditions hold:
(a) if $\mathcal{F}(C)$ is an atom of the form $t=\mbox{`ct'}$ where $t$ is a term for temporal attributes of items, $\mbox{`ct'}$ is a temporal constant in the temporal domain with some granularity and $\mbox{`ct'}=t_m$;
(b) if $\mathcal{F}(C)$ is an atom of the form   $f_n(t, \mbox{`ct1'}, \mbox{`ct2'},...,\mbox{`ct}_{n-1}\mbox{'})=\mbox{`ct'},$  where $t$ is a temporal attribute, $\mbox{`ct'}$
 is a temporal constant, and  $f_n(t_m, \mbox{`ct1'}, \mbox{`ct2'},...,\mbox{`ct}_{n-1}\mbox{'})=\mbox{`ct'}$ ;
  (c) if $\mathcal{F}(C)$ does not contain a temporal attribute;
 (d) if $\mathcal{F}(C)$ is a formula  $F1 \land F2$ and F1  and F2 are satisfied by $\mu$.\qed
 \end{definition}

%
%

\begin{definition}{\rm [Total Matching of a Sequential Expression]}
\label{def:totalmatching}
Given a sequential expression SE=$SE_1$.$SE_2$...$SE_k$ of length k, and
a \emph{normalized ToI} with schema $(OID, t,  ID),$
we say that an object identified by $OID_m$ \emph{totally matches} SE,
if there exists $k$ different tuples $\mu_1,\ldots,\mu_k$ in nToI, of the form
$\mu_1=\langle OID_m, t_1, ID_1\rangle,$ $\mu_2=\langle OID_m, t_2, ID_2\rangle$,...,$\mu_k=\langle OID_m, t_k, ID_k\rangle$, and there is  at least one list $L_p=(eco_1, eco_2, ... eco_k), L_p~\in~\mathcal{S}_{L_{k}}(SE),$ where
the following conditions hold: (a) $t_i~\in~\mathit{Interval}(eco_i),~\forall~i,i=1..k;$ (b) $ID_i$ is the identifier of the encoded category occurrence $(eco_i),~\forall~i,i=1..k;$ (c) $SE_i$ is temporally satisfied by $\mu_i,~\forall~i,=1..k.$
We denote each  $L_p$ a \emph{list of interest} for SE.\qed
\end{definition}

\begin{property}
Given a sequential expression SE=$SE_1$.$SE_2$...$SE_k$ of length k, and
a \emph{normalized ToI} $\mathcal{T}$ with schema $(OID, t,  ID).$ If an object $OID_j$ in  $\mathcal{T}$
 does not temporally match SE, then $OID_j$ cannot \emph{totally match} SE.\qed
\end{property}

\begin{property}
Given a sequential expression SE=$SE_1$.$SE_2$...$SE_k$ of length k, and
a \emph{normalized ToI} $\mathcal{T}$ with schema $(OID, t,  ID),$
such that there is an object $OID_j$ in  $\mathcal{T}$
 that \emph{totally matches} SE, then $OID_j$  \emph{temporally matches} SE. \qed
\end{property}

\begin{example}
 Object $Session_{1}$ in  Example \ref{ex:temporalmatch},  totally matches SE, using the
 second and third tuples, together with list $L_1$. On the other hand,  $Session_{2}$ does not totally match SE,
 since it does not temporally match the expression. Finally,   $Session_{3}$ temporally matches SE,
  but it does not totally match it,  because   $L_2$ does not satisfy the second condition in Definition \ref{def:totalmatching}.\qed
\end{example}


\begin{definition}[Temporal Support of SE]
\label{def:temporalsupport1}
The \emph{temporal support}  of a sequential expression SE, denoted $\mathcal{T}_s(SE),$ is the quotient between the number of different objects that \emph{totally match SE} and the number of different objects that \emph{temporally match SE}, if the latter is different to zero. Otherwise $\mathcal{T}_s(SE)=0$. \qed
\end{definition}

Definition \ref{def:temporalsupport1} formalizes the intuition behind the concept of temporal support, namely, counting only
the sequences that could have potentially generated a matching sequence, given the temporal availability of the category occurrences
to which an item in a sequence belong (these sequences are the ones that \emph{temporally match} a SE).
Classic support count, instead, considers the whole number of sequences in the database.

\begin{example}
In the example above $\mathcal{T}_s([ID=\mbox{`P'}].[\mathit{filter}=\mbox{`M'}])=0.5.$  The object $Session_2$ is not considered in the support count because when the user clicked the Web page, it had not the possibility of selecting  pages that satisfy the constraint. \qed
\end{example}


\section{Temporal Support of Regular Expressions}
\label{sec:querylanguage}

Having defined the temporal support of a sequential expression, we now move on to
the general problem, i.e., defining the same concept for an RE.
 The data model defined in Section \ref{sec:datamodel}, and the theory developed in Section
  \ref{sec:simplelanguage},  allows us to define a language based on RE over  constraints, that supports categorical attributes.
 We start with a simple example.  We wish to restrict the result of an SPM  algorithm to
 the sequences that match the following expressions: (a)  $SE_1$=[keyword=`Games']; (b)
   $SE_2$ = [keyword=`Games'].[filter=`']; (c) $SE_3$ = [keyword=`Games'].[filter=`'].[filter=`']. For   each
   $SE_i,~i>3$, a condition [filter=`'] is added.
 We are also  interested in computing the  temporal support of these sequential expressions. Instead of computing
 each support in a separate fashion, we may want to  summarize these sequences in a single  RE, namely:
 $[\mathit{keyword}=\mbox{`Games'}].([filter=\mbox{`'}])^*$. 


\begin{definition}{\rm[R.E. over constraints]}
\label{def:reconst}
A \emph{regular expression} over the constraints of Definition \ref{def:constraint},
 is an expression generated by the grammar

\[ E \longleftarrow   C  \mid E|E \mid  E?
    \mid  E^{*} \mid E^{+} \mid E.E \mid E \mid\epsilon \]

where $C$ is a constraint, $\epsilon$ is the symbol representing the empty expression, $\mbox{`}~|\mbox{'}$ means disjunction,
$\mbox{`}~.\mbox{'}$ means concatenation,  $\mbox{`}~?\mbox{'}$   ``zero or one occurrence'', $\mbox{`}~+\mbox{'}$ ``one or more occurrences'',
 and $\mbox{`}~*\mbox{'}$   ``zero or more occurrences''. The precedence is the usual one.\qed
\end{definition}

\begin{property}
\label{prop:regexpe}
Let $\mathcal{L}$ be the set of sequential expressions $SE_i$ produced   by a RE $\mathcal{R},$ generated by the grammar of Definition \ref{def:reconst}.
There is also a normalized ToI with schema $(OID,t,ID).$
If an object $O_i$ in the nToI temporally or totally matches  any SE  in  $\mathcal{L},$ $O_i$  matches $\mathcal{R}$,
temporally or totally, respectively. \qed
\end{property}

Property \ref{prop:regexpe} follows from observing that
 [keyword=`Games'].[filter=`']* could be written:
[keyword=`Games'] $\mid$ ([keyword=`Games'].[filter=`']) $\mid$ ([keyword= `Games'].[filter=`'].[filter=`']) $\mid$
([keyword=`Games'].[filter=`'].[filter=`'].[filter=`']) $\mid \ldots.$

 Reasoning along the same lines, since a regular expression over an alphabet (in our case, constraints)
 denotes the language that is recognized by a Deterministic Finite Automata (DFA), there exists a (possible infinite)
   set of strings over the alphabet that this DFA accepts. Each of these strings (actually, strings composed of constraints)
  matches  our definition of SE. Then, we extend our previous  definition of temporal and total matching of
  SE, to RE, as  follows.

\begin{definition}[Temporal Matching of a RE]
\label{def:temporalmatchingRE}
Given a regular expression $\mathcal{R}$ generated by the grammar of Definition \ref{def:reconst}, and the  DFA $\mathcal{A}_\mathcal{R}$ that accepts $\mathcal{R}.$ There is also a  normalized ToI  with schema $(OID, t,  ID).$
We say that $OID_m$ \emph{temporally matches} $\mathcal{R},$ if there exists some $n \in \mathbb{N}$ such
that there exists at least one string of length $n$ accepted by $\mathcal{A}_\mathcal{R},$
and $OID_m$ \emph{temporally matches} this string. \qed
\end{definition}

\begin{definition}[Total Matching of a RE]
\label{def:totalmatchingRE}
Given a regular expression $\mathcal{R}$ generated by the grammar of Definition \ref{def:reconst}, and the  DFA $\mathcal{A}_\mathcal{R}$ that accepts $\mathcal{R}.$ There is also a  normalized ToI  with schema $(OID, t,  ID).$
We say that $OID_m$ \emph{totally matches} RE, if there exists some $n \in \mathbb{N}$
such that there  is at least one string of length n accepted by $\mathcal{A}_\mathcal{R}$ and
$OID_m$ \emph{totally matches} this string. \qed
\end{definition}

\begin{definition}[Temporal Support of a RE]
\label{def:temporalsupport2}
The \emph{temporal support} of a regular expression $\mathcal{R},$ denoted $\mathcal{T}_r(\mathcal{R}),$
is the quotient between the
number of different objects that totally match $\mathcal{R}$ and the number of different objects
that temporally match $\mathcal{R}$, if the latter is different to zero. Otherwise $\mathcal{T}_r(\mathcal{R})=0$. \qed
\end{definition}



 We  use the example above to show how  \emph{sequential expressions} are summarized using  the language of
  Definition \ref{def:reconst}.
 We use the category occurrences and the  nToI shown in Figures \ref{fig:eco-example} and  \ref{fig:ntoi-example},
 respectively.
 We first  apply Definition \ref{def:constraintsatisf} in order to  check  satisfability of the constraints in  the expressions $SE_1$ through $SE_3.$  ECOs  $eco_{C1}$, $eco_{C2}$ and $eco_{C3}$ from Figure \ref{fig:eco-example}  satisfy the constraint [keyword=`Games']. Analogously,  the constraint [filter=`'] is satisfied by  $eco_{A1}$, $eco_{C2}$, $eco_{M1}$ and $eco_{P1}$.
  Next, for each  SE, we  check  satisfability  applying Definition \ref{def:SEsatisfability}.
  For  $SE_1$=[keyword=`Games'] we obtain $\mathcal{S}_{L_{1}}(SE)= \{L_{1}=\{eco_{C2}\}$, $L_{2}= \{eco_{C3}\},$ $L_{3}= \{eco_{M1}\}\}$. For $SE_2$=[keyword=`Games'].[filter=`'] we have  $\mathcal{S}_{L_{2}}(SE)= \{L_{1}=\{eco_{C2},eco_{C2}\}$,
  $L_{2}= \{eco_{C2},eco_{M1}\},$  $L_{3}= \{eco_{C2},eco_{P1}\},$  $L_{4}= \{eco_{C3},eco_{M1}\},$  $L_{5}= \{eco_{M1},eco_{M1}\}.$ Note
  that, for example, the list $\{eco_{C2},eco_{A1}\}$ does not satisfy $SE_2$ because $eco_{C2}$ \emph{follows} $eco_{A1}$. For  $SE_3$= [keyword=`Games'].[filter=`'].[filter=`'], we have
   $\mathcal{S}_{L_{3}}(SE)=  \{L_{1}=\{eco_{C2},eco_{C2},$  $eco_{C2}\}$,  $\{L_{2}=\{eco_{C2},eco_{C2},eco_{M1}\}$,
    $\{L_{3}=\{eco_{C2},eco_{C2},eco_{P1}\}$, $L_{4}= \{eco_{C2},$  $eco_{M1}, eco_{M1}\}$,
    $L_{5}= \{ eco_{C2}, eco_{P1},$ $eco_{C2}\},$  $L_{6}= \{eco_{C2}, eco_{P1},$ $eco_{M1}\},$
    $L_{7}= \{ eco_{C2}, eco_{P1},$ $eco_{P1}\},$ $L_{8}= \{ eco_{C3}, eco_{M1},$ $eco_{M1}\},$ $L_{9}= \{ eco_{M1}, eco_{M1},$ $eco_{M1}\}$.
     Also here, many lists are discarded. For instance, $\{eco_{M1}, eco_{M1}, eco_{A1}\}$ does not satisfy $SE_3$ because $eco_{M1}$ \emph{follows} $eco_{A1}$.

 Now, we can compute  the temporal support of each SE,  applying Definitions \ref{def:temporalmatching} through \ref{def:temporalsupport1}.
 For $SE_1,$ from  the third tuple in $Session_1,$  and  $L_{2}= \{eco_{C3}\},$ we conclude that $Session_1$ totally
 (and hence, temporally) matches  $SE_1$.
  From the first tuple in $Session_2$ and  $L_{1}= \{eco_{C2}\},$  $Session_2$   totally  matches $SE_1$.  From the first or third tuples in $Session_3,$ and    $L_{2}= \{eco_{C3}\},$  $Session_3$ totally matches $SE_1$.
  Finally,  the temporal support of $SE_1$ is $3/3=1.$

   In a similar way,  we can conclude that
  the support of $SE_2$ and $SE_3$ are, respectively, $1$  and $1/2.$
    Since no session has four tuples, it is not necessary to analyze a  sequential expression of length four, like for instance [keyword=`Games'].$  \\ $[filter=`'].[filter=`'].[filter=`'].


%
%
%
%
%
%
%
%
%
%
%
%
%
%
%
%
%
%

\section{Future Work}
\label{sec:conclu}
  We expect to extend our work in two ways. On the one hand, the theoretical
  framework introduced here allows to think in a more general definition of support,
  with different semantics (not only temporal), that may   enhance  current data mining  tools.
  On the other hand, we will
  develop an optimized  implementation of the algorithm that can support massive amounts
  of data.  
  
\bibliographystyle{plain}





\end{document}